\documentclass[letterpaper, 10 pt, conference]{ieeeconf}
\IEEEoverridecommandlockouts
\makeatletter

\let\proof\@undefined
\let\endproof\@undefined
\makeatother

\usepackage{cite}
\usepackage{amsmath,amssymb,amsfonts}
\usepackage{algorithm}
\usepackage{algorithmicx}
\usepackage{algpseudocode}
\usepackage{graphicx} 
\usepackage{textcomp}
\usepackage{xcolor}
\usepackage{caption}
\usepackage{array}
\usepackage{float}
\usepackage{dsfont}
\usepackage{amsthm}
\newcommand{\subparagraph}{}
\usepackage[compact]{titlesec}
\usepackage{xspace}

\usepackage[textwidth=1.5in]{todonotes}

\renewcommand{\thetable}{\arabic{table}}
\def\BibTeX{{\rm B\kern-.05em{\sc i\kern-.025em b}\kern-.08em
    T\kern-.1667em\lower.7ex\hbox{E}\kern-.125emX}}    
\usepackage{etoolbox}
\usepackage{url}
\newtheorem{definition}{Definition}
\newtheorem{remark}{Remark}
\newtheorem{theorem}{Theorem}
\newtheorem{fact}{Fact}
\newtheorem{proposition}{Proposition}

\DeclareMathOperator{\vect}{vec}

\newcommand{\MATLAB}{\textsc{Matlab}\xspace}

\newfloat{algorithm}{t}{lop}

\begin{document}
\title{\textbf{\huge{Verification and Control for Finite-Time Safety of Stochastic Systems via Barrier Functions}}
\author{Cesar Santoyo, Maxence Dutreix, and Samuel Coogan \vspace{-10mm}
\thanks{C. Santoyo ({\tt \small csantoyo@gatech.edu}) and M. Dutreix ({\tt \small maxdutreix@gatech.edu}) are with the School of Electrical \& Computer Engineering, Georgia Institute of Technology, Atlanta, GA., 30318, USA.}
\thanks{S. Coogan ({\tt \small sam.coogan@gatech.edu}) is with the School of Electrical \& Computer Engineering and the School of Civil and Environmental Engineering, Georgia Institute of Technology, Atlanta, GA., 30318, USA.}
\thanks{This work was partially supported by NSF under Grant \#1749357. C. Santoyo was supported by the NSF Graduate Research Fellowship Program under Grant No. DGE-1650044.}}
}
\maketitle
\begin{abstract}
  This paper studies the problem of enforcing safety of a stochastic dynamical system over a finite time horizon. We use stochastic barrier functions as a means to quantify the probability that a system exits a given safe region of the state space in finite time. A barrier certificate condition that bounds the infinitesimal generator of the system, and hence bounds the expected value of the barrier function over the time horizon, is recast as a sum-of-squares optimization problem for efficient numerical computation. Unlike prior works, the proposed certificate condition includes  a state-dependent bound on the infinitesimal generator, allowing for tighter probability bounds.  Moreover, for stochastic systems for which the drift dynamics are affine-in-control, we propose a method for synthesizing polynomial state feedback controllers that achieve a specified probability of safety. Two case studies are presented that benchmark and illustrate the performance of our method.% failure probability upper bound against the true values.
  
%In this paper, we address both the problem of stochastic system safety verification on a finite-time horizon and minimum energy control policy search by introducing the novel concept of \textit{stochastic control barrier functions}. In the deterministic setting, barrier functions are used to certify a system will remain within a desired region; however, stochastic safety verification requires a quantification of the probability a system exits the safe region. Our approach relaxes the constraint on the evolution of the expected value of the barrier function. We present a novel, state-dependent approach for finite-time horizon verification. Furthermore, we address the problem of controller synthesis given a desired criterion on a system's failure probability. Lastly, we present our results regarding a low energy control policy at particular noise levels for a desired failure probability criterion.
\end{abstract}
\section{Introduction}
Reliance on complex, safety-critical systems is increasing, which has made safety verification of such systems of utmost importance. For example, environments populated by both humans and autonomous systems (e.g. fulfillment centers, autonomous vehicles, and healthcare) require rigorous safety verification to ensure desired behavior is achieved. From a practical standpoint, safety verification can translate directly to ensuring qualitative guidelines such as collision avoidance are maintained. Safety-critical systems are often analyzed in a purely deterministic framework, however, many real-world applications are subject to stochastic disturbances and are better modeled as stochastic systems.  \par 
A common approach to safety verification in deterministic systems is via \textit{barrier functions} which provide Lyapunov-like guarantees regarding system behavior. The existence of a barrier function which satisfies a \textit{barrier certificate} can often be enough to certify the safe operation of a system \cite{PrajnaStoch2007}. %within a specified time horizon. Some of the earliest work on barrier functions in the controls field was shown in.
%Since then,
Recent work has modified and improved the deterministic form of barrier functions and expanded their application. In particular, control barrier functions have been introduced to guaranteed safety in control affine systems \cite{wieland2007constructive, AmesAaronD.2017CBFB}.  This is demonstrated in applications for cruise control \cite{AmesAaronD.2017CBFB, ames2014control}, collision avoidance in robotic swarms \cite{WangLi2017SBCf}, and walking robots \cite{Hsu:2015uo}, and has recently been extended to allow for input-to-state safe control barrier functions \cite{KolathayaShishir2018ISWC} and to guarantee %Additionally, guarantees of
finite-time convergence to a safe region %has been proven utilizing control barrier functions,
\cite{AnqiLi}. \par 
In the stochastic setting, safety verification via barrier certificates for infinite time horizons was introduced in \cite{PrajnaStoch2007} alongside the deterministic counterpart. The work presented in \cite{PrajnaStoch2007} provides a framework for bounding the probability a system will exit a safe region based on a non-negative barrier function defined on the system state space. In this approach, the probability is directly correlated with the set of initial conditions. However, this approach can be overly restrictive because it requires the infinitesimal generator, which dictates the expected value evolution of a stochastic process, to be non-positive; i.e., the barrier function is restricted to be a \textit{supermartingale}. \par
The paper \cite{SteinhardtJacob2012Frvo} relaxes this condition and instead provides a barrier certificate that only requires the infinitesimal generator of the barrier process to be upper-bounded by a constant. Such processes are called \textit{c-martingales} and allow the expected value of the barrier function to increase over time. This approach results in a safety probability bound for finite time horizons. Recent work in \cite{Pushpak_Sadegh_Zamani_2018} leverages c-martingales for temporal logic verification of discrete-time systems.\par

The present paper also studies the problem of verifying safety of stochastic systems on finite time horizons, and the contributions are as follows. First, we build on the approaches proposed in \cite{PrajnaStoch2007, SteinhardtJacob2012Frvo} and propose a barrier certificate constraint that imposes a state-dependent bound on the infinitesimal generator. This bound was originally proposed and studied by Kushner in \cite{Kushnerstochstabtext, KushnerH.1966Ftss}. The proposed barrier certificate allows the expected value of the barrier to increase and covers the c-martingale condition of \cite{SteinhardtJacob2012Frvo} as a special case. However, our formulation also accounts for the system dynamics in the infinitesimal generator constraint. This allows for probability bounds that are no worse than the c-martingale condition, and in many cases, especially with high noise levels, provides better probability bounds. \par 
Second, as in \cite{PrajnaStoch2007, SteinhardtJacob2012Frvo}, we compute barrier functions using \textit{sum-of-squares} (SOS) optimization. Like in \cite{PrajnaStoch2007}, but unlike \cite{SteinhardtJacob2012Frvo}, we utilize polynomial barrier functions. This provides a simpler formulation of the probability of failure on a finite time horizon when compared to the approach in \cite{SteinhardtJacob2012Frvo} which uses exponential barrier functions and, empirically, provides tighter probability bounds. \par 
Third, we extend our formulation to allow for control inputs and provide a method for synthesizing a safe controller. In particular, we consider affine-in-control systems and the proposed approach searches for a polynomial state feedback controller which ensures a system's failure probability achieves a predetermined criterion via a \emph{stochastic control barrier function}. \par 
This paper is organized as follows: Section II covers the background information of stochastic differential equations, barrier functions and SOS optimization. Section III covers the problem which we are solving in detail. Section IV highlights the methodology we utilize to solve the SOS optimization and stochastic control problem.  Section V and Section VI present numerical case studies which illustrate our results and conclusions, respectively. \par 
\section{Preliminaries}
In this section we first introduce our state space definitions as well as background information regarding stochastic processes, barrier functions, and SOS polynomials.
\subsection{Stochastic Process}
Consider a complete probability space ($\Omega, \mathcal{F}, P$) and a standard Wiener process, $w(t)$ in $\mathbb{R}^m$. We consider stochastic processes $x(t)$ satisfying a stochastic differential equation of the form
\begin{equation} \label{stochasticdiffeq}
	dx = F(x)dt + \sigma(x)dw.
\end{equation}
 The compact set $\mathcal{X} \subset \mathbb{R}^n$ is the system state space, $F: \mathcal{X} \rightarrow \mathbb{R}^n$ is the drift rate and $\sigma: \mathcal{X} \rightarrow \mathbb{R}^{n \times m}$ is the diffusion term. We assume the functions $F(x)$ and $\sigma(x)$ are Lipschitz continuous. The stochastic process $x$ is a right continuous strong Markov process \cite{ksendalB.K.BerntKarsten1998Sde:}. We now introduce the infinitesimal generator, which extends the usual definition of a time derivative to instead consider the expectation of a function of a random process.
\begin{definition}
Let $x$ be a stochastic process in $\mathbb{R}^n$. The infinitesimal generator $\mathcal{A}$ of $x$ acts on functions of the state space and is defined as
\begin{equation}
\nonumber
\mathcal{A}B(x) = \lim\limits_{t \downarrow 0} \frac{\mathbb{E}[B(x) | x_0] - B(x_0)}{t}
\end{equation}
where $B: \mathcal{X} \rightarrow \mathbb{R}$ such that the limit exists for all $x_0$. 
\end{definition} \par 
In particular, the infinitesimal generator for any process as in (\ref{stochasticdiffeq}) is of the form shown in Fact \ref{generatorfact}. 
\begin{fact}[{Ch. 7, Theorem 7.3.3 of \cite{ksendalB.K.BerntKarsten1998Sde:}}]
	Let $x$ be a stochastic process satisfying (\ref{stochasticdiffeq}), then the infinitesimal generator $\mathcal{A}$ of some twice differentiable function $B(x)$ is given by
\begin{equation}
		\nonumber
		\mathcal{A}B(x) =  \sum_{i = 1}^{n}F_i(x)\frac{\partial B }{\partial x_i} + \frac{1}{2}\sum_{i = 1}^{n}\sum_{j = 1}^{n}\bigg(\sigma(x) \sigma^T(x)\bigg)_{i,j}\frac{\partial^2B}{\partial x_i \partial x_j}.
	\end{equation}
	\label{generatorfact}
\end{fact}
\par The stochastic process $x$ is not guaranteed to lie in $\mathcal{X}$ at all times which leads us to define the stopped process $\tilde{x}$.
\begin{definition}
	Suppose that $\tau$ is the first time of exit of $x$ from the open set Int($\mathcal{X}$). Then the stopped process $\tilde{x}$ is defined by \cite{PrajnaStoch2007}
	\begin{equation} \nonumber
	\tilde{x}(t) = 
		\begin{cases} 
		x(t) & \text{for} \ t \leq \tau \\
		x(\tau) & \text{for} \ t \geq \tau.
		\end{cases}
	\end{equation}
\end{definition}
It is worth noting that the stopped process inherits the same strong Markovian property of $x$ and shares the same infinitesimal generator \cite{Kushnerstochstabtext}.
\subsection{Barrier Functions}
Consider an unsafe region of the state space $\mathcal{X}_u \subseteq \mathcal{X}$ and a set of initial conditions $\mathcal{X}_0 \subseteq \mathcal{X} \setminus \mathcal{X}_u$. In a similar spirit to Lyapunov functions, barrier functions are utilized as a means of guaranteeing a desired behavior on some region of a system's domain defined as a sub-level set (or super-level set) of the barriers. In that regard, stochastic barrier functions have been introduced to upper bound the probability of exiting a safe region over an infinite time-horizon.
\begin{proposition}[Theorem 15 from \cite{PrajnaStoch2007}]
	Given a stochastic differential equation of the form of (\ref{stochasticdiffeq}) and the sets $\mathcal{X}$, $\mathcal{X}_0$, and $\mathcal{X}_u$ with $f(x)$ and $\sigma(x)$ locally Lipschitz continuous, consider the stopped process $\tilde{x}$. Suppose there exists a twice differentiable function $B$ such that
	\begin{equation}
		B(x) \leq \gamma \ \forall x \in \mathcal{X}_0
	\end{equation}
		\begin{equation}
	B(x) \geq 1 \ \forall x \in \mathcal{X}_u
	\end{equation}
	\begin{equation}
	B(x) \geq 0 \ \forall x \in \mathcal{X}
	\end{equation}
	\begin{equation} \label{supercond}
	\frac{\partial B}{\partial x}f(x) + \frac{1}{2}\text{Trace}\bigg(\sigma ^T(x) \frac{\partial^2B}{\partial x^2}\sigma (x) \bigg) \leq 0 \ \ \forall x \in \mathcal{X}.
	\end{equation}
	Then, the probability of the system entering the unsafe region of the state space is bounded by
	\begin{equation}
          \label{eq:1}
		P \{\tilde{x}(t)\in \mathcal{X}_u \text{ for some } t\geq 0 \} \leq B(x_0) \leq \gamma
              \end{equation}
              where $x_0\in \mathcal{X}_0$ is the initial state of the system.
            \end{proposition}
This theorem provides a powerful means of bounding the probability of failure of a stochastic process on an infinite time horizon. However,  we note that the inequality condition (\ref{supercond}), also referred to as the barrier certificate, imposes that $B(x)$ is a supermartingale. This inequality enforces that the expectation of the barrier function decreases at all points of $\mathcal{X}$. In practice, this is often overly restrictive on the system dynamics. For example, it has been shown that no supermartingale exists on a bounded set where the system's noise does not vanish \cite{Kushnerstochstabtext}. In Section III, we present a relaxed version of this theorem with its respective probability bounds for finite-time horizons.
\subsection{Sum-of-Squares}
\begin{definition}
	\label{sumofsquaresdefinition}
	Define $\mathbb{R}[x]$ as the set of all polynomials in $x\in\mathbb{R}^n$. Then
	\begin{equation}
		\nonumber
		\Sigma[x] \triangleq \bigg\{   s(x) \in \mathbb{R}[x] : s(x) = \sum_{i=1}^{m} g_i(x)^2, g_i(x) \in \mathbb{R}[x] \bigg\}
	\end{equation}
	is the set of SOS polynomials. It is noted that if $s(x) \in \Sigma[x]$ then $s(x) \geq 0$ $\forall$ $x$.
\end{definition}
\begin{definition}
	Given $p_i(x) \in \mathbb{R}[x]$ for $i = 0, \ldots, m$, the problem of finding $q_i(x) \in \Sigma[x]$ for $i = 1, \ldots, \hat{m}$ and $q_i(x)\in\mathbb{R}[x]$ for $i=\hat{m}+1,\ldots, m$ such that
	\begin{equation}
		\nonumber
		p_0(x) + \sum_{i = 1}^{m}p_i(x)q_i(x) \in \Sigma[x]
	\end{equation}
	is a sum-of-squares program (SOSP). SOSPs can be efficiently converted to semidefinite programs using tools such as SOSTOOLS \cite{sostools}.
      \end{definition}
\section{Problem Formulation}
The problem we address is: how do we create a bound on the probability a stochastic system of form (\ref{stochasticdiffeq}) exits a safe region during a finite-time horizon? \par 
\textbf{Objectives:} First, our goal in this paper is to relax the supermartingale condition on the barrier certificate in (\ref{supercond}) similar to what is shown in \cite{SteinhardtJacob2012Frvo}. Second, based on that relaxation, we aim to derive a state-feedback controller ensuring a user-specified upper bound on the probability of exiting a safe region in the state space. \par 

Consider the stochastic process $x$ which satisfies the stochastic differential equation
\begin{equation}
\label{controlstochdiff}
dx = (f(x) + g(x)u(x) )dt + \sigma(x)dw
\end{equation}
%where $F(x) = f(x) + g(x)u(x)$.
%We have
where $f:\mathcal{X} \rightarrow \mathbb{R}^n $, $g:\mathcal{X} \rightarrow \mathbb{R}^{n \times k} $, $\sigma:\mathcal{X} \rightarrow \mathbb{R}^{n \times m}$ and  $w$ is a $m$-dimensional Wiener process. Additionally, $u: \mathcal{X} \rightarrow \mathbb{R}^k$ where $u$ is a state dependent control input. We define $F(x) = f(x) + g(x)u(x)$.\par 
Now, we relax the supermartingale condition shown in (\ref{supercond}). The following theorem is an immediate corollary of Chapter 3, Theorem 1 in \cite{Kushnerstochstabtext}. \vspace{-1mm}
\begin{theorem} \label{alphatheorem}
	Given the stochastic differential equation shown in (\ref{controlstochdiff}) and the sets $\mathcal{X} \subset \mathbb{R}^n$, $\mathcal{X}_u \subseteq \mathcal{X}, \mathcal{X}_0 \subseteq \mathcal{X}\setminus \mathcal{X}_u$ with $F(x) = f(x) + g(x)u(x)$ and $\sigma(x)$ locally Lipschitz continuous, where $u(x)$ is some feedback control strategy. Consider the stopped process $\tilde{x}$. Suppose there exists a twice differentiable function $B$ such that
	\begin{equation}
	B(x) \leq \gamma \ \forall x \in \mathcal{X}_0
	\end{equation}
	\begin{equation}
	B(x) \geq 1 \ \forall x \in \mathcal{X}_u
	\end{equation}
	\begin{equation}
	B(x) \geq 0 \ \forall x \in \mathcal{X}
	\end{equation}
	\begin{equation} \label{supmartrelax}
	\small \small
	\frac{\partial B}{\partial x}F(x) + \frac{1}{2}\text{Trace}\bigg(\sigma ^T(x) \frac{\partial^2B}{\partial x^2}\sigma (x) \bigg) \leq -\alpha B(x) + \beta \ \ \ \forall x \in \mathcal{X}\setminus \mathcal{X}_u
	\end{equation}
	for some $\alpha \geq 0$, $\beta \geq 0$ and $\gamma \in [0, 1)$. Define
        \begin{align}
          \label{eq:2}
          \rho_{u}&:=P \{\tilde{x}(t)\in \mathcal{X}_u \text{ for some } 0\leq t\leq T \}. 
          %\label{eq:3}
                  %&\leq	
        \end{align}
Then 
\begin{itemize}
\item If $\alpha > 0$ and $\frac{\beta}{\alpha} \leq 1$,
	\begin{equation}
	\label{bound1}
%	P\left\{\sup_{0\leq t \leq T} B(x) \geq 1 \right\}
\rho_u\leq P\left\{\sup_{0\leq t \leq T} B(\tilde{x}) \geq 1 \right\}
        \leq 1 - \bigg(1 - B(x_0) \bigg)e^{-\beta T}.
	\end{equation}
\item If $\alpha > 0$ and $\frac{\beta}{\alpha} \geq 1$,
	\begin{equation}
	\label{bound2}
\rho_u\leq P\left\{\sup_{0\leq t \leq T} B(\tilde{x}) \geq 1 \right\}\leq \frac{B(x_0) + (e^{\beta T} - 1)\frac{\beta}{\alpha}}{e^{\beta T}}.
	\end{equation}
\item If $\alpha = 0$,
    \begin{equation}
	\label{bound3}
\rho_u\leq P\left\{\sup_{0\leq t \leq T} B(\tilde{x}) \geq 1 \right\} \leq B(x_0) + \beta T.
      \end{equation}
    \end{itemize}
  \end{theorem} \par
%   \SC[inline]{I reformatted the bullets again. Please see if you like it.}
%   \CS[inline]{Looks good.}
 %   \SC[inline]{I rewrote this as a bulleted list. Please see if you agree.}
%    \CS[inline]{Looks good.}
    % \SC[inline]{These probablity bounds aren't written in terms of the unsafe set. I think it woudl be better to do so.}
    % \CS[inline]{Written as $P \{\tilde{x}(t)\in \mathcal{X}_u \text{ for some } t\geq 0 \}$? If so, I can do that.}
    % \SC[inline]{What do you think of the above? Also, should \eqref{eq:3} be $B(\tilde{x})$?}
    % \CS[inline]{Looks good. Yes. Fixed.}

The bound shown in (\ref{bound3}) is characterized in \cite{Pushpak_Sadegh_Zamani_2018} and \cite{SteinhardtJacob2012Frvo} as the upper bound on the probability of being unsafe for a c-martingale. \par
If $B(x)$ satisfies the conditions of Theorem \ref{alphatheorem}, then $B(x)$ is called a \emph{stochastic control barrier function} for a given control policy $u(x)$. Relaxing the supermartingale condition on the infinitesimal generator in the fashion of Theorem \ref{alphatheorem} gives three case-dependent finite time probability bounds on a system's likelihood of entering an unsafe region in the form of (\ref{bound1}), (\ref{bound2}), and (\ref{bound3}). \vspace{-1mm}
\begin{remark}
  \label{rem:1}
If the initial state $x_0$ is not known exactly but only known to lie within $\mathcal{X}_0$, then $\gamma$ can be substituted for $B(x_0)$ in the probability bounds in Theorem \ref{alphatheorem}. This provides an upper bound on the probability of failure over the entire set of initial conditions rather than on a particular initial point in $\mathcal{X}_0$.  \vspace{-2mm}
\end{remark}
\section{Methodology} \vspace{-1mm}
In this section we present our approach to construct the stochastic control barrier functions based on the problem formulation of Section II. First, we adapt the constraints given in Theorem \ref{alphatheorem} to be formulated as an SOSP. Second, we cover the algorithms which construct barrier functions and present our method for computing a low-energy control policy $u(x)$. \vspace{-1mm} \par
\subsection{SOS Formulation for Safety Verification}\vspace{-1mm}
\begin{theorem} \label{SOSPTheorem}
    Consider a system of the form of (\ref{controlstochdiff}) and the sets $\mathcal{X}$, $\mathcal{X}_0$, and $\mathcal{X}_u$ and assume these sets can be described as $\mathcal{X} = \{x\in \mathbb{R}^n : s_{\mathcal{X}}(x) \geq 0 \}$, $\mathcal{X}_0 = \{x\in \mathbb{R}^n : s_{\mathcal{X}_o}(x) \geq 0 \}$, and $\mathcal{X}_u = \{x\in \mathbb{R}^n : s_{\mathcal{X}_u}(x) \geq 0 \}$ for some polynomials $s_{\mathcal{X}}$, $s_{\mathcal{X}_o}$, and $s_{\mathcal{X}_u}$. Suppose there exists a polynomial $B(x)$, a polynomial $u(x)$, and SOS polynomials $\lambda_{\mathcal{X}}(x)$, $\lambda_{\mathcal{X}_o}(x)$, and $\lambda_{\mathcal{X}_u}(x)$ that satisfy the following
    \begin{equation} \nonumber
        B(x) - \lambda_{\mathcal{X}}(x)s_{\mathcal{X}}(x) \in \Sigma[x]
    \end{equation}
    \begin{equation} \nonumber
        B(x) - \lambda_{\mathcal{X}_u}(x)s_{\mathcal{X}_u}(x) - 1 \in \Sigma[x]
    \end{equation}
        \begin{equation} \nonumber
        -B(x) - \lambda_{\mathcal{X}_o}(x)s_{\mathcal{X}_o}(x) + \gamma \in \Sigma[x]
    \end{equation}
    \begin{align} \nonumber 
        -\frac{\partial B(x)}{\partial x}F(x) - \alpha B(x) + \beta + \lambda_{\mathcal{X}_u}(x)s_{\mathcal{X}_u}(x)\\  \nonumber  - \lambda_{\mathcal{X}}(x)s_{\mathcal{X}}(x) \in \Sigma[x] 
    \end{align}
    where $F(x) = f(x) + g(x)u(x)$. Then, the probability of failure, depending on the values of $\alpha$ and $\beta$, is defined by (\ref{bound1}), (\ref{bound2}) or (\ref{bound3}).
\end{theorem} \vspace{-1mm}
%   \SC[inline]{Why do we have two different functions $ s_{\mathcal{X}_u}(x) $ and $s_{{\mathcal{X} \setminus \mathcal{X}_u}}(x)$?}
%   \CS[inline]{The polynomial $s_{{\mathcal{X} \setminus \mathcal{X}_u}}(x)$ encodes the state space without the unsafe set. We discussed in one of our meetings enforcing the relaxed supermartingale condition only within the safe region, $\mathcal{X} \setminus \mathcal{X}_u$. $s_{\mathcal{X}_u}$ is a polynomial encoding the unsafe region.}
% \SC[inline]{I guess I'm still confused if there's a relationship between the two. Is one just the negative of the other?}
%   \CS[inline]{Oh, I see what you mean. Yes, that is true. They are negatives of the other.}
%   \SC[inline]{Ok, can we write as $+ \lambda_{\mathcal{X}\setminus \mathcal{X}_u}(x)s_{\mathcal{X}_u}(x)$ instead then?}
We omit the proof due to space constraints, but the proof follows the general approach for relaxing set constraints to SOS programs using the \emph{Positivstellensatz} condition; see the documentation of \cite{sostools} for details. %\vspace{10ex}% constructing SOS programs for  for SOS programs of the same form. %\QEDA
% \begin{proof}
% \end{proof}
    \par 
    \subsection{Barrier Function Numerical Procedure}
	Next, we present an algorithmic solution to this problem. Algorithm \ref{Bxalgo} computes the barrier function $B(x)$ used to quantify an upper bound on the failure probability. The input values $l_{\alpha}, u_{\alpha}, \sigma, u(x), n_B$ are the lower $\alpha$ range value, upper $\alpha$ range value, diffusion term, control polynomial, and the order of the $B(x)$ polynomial, respectively. Our algorithm performs a grid search over a range of scalar $\alpha$ with value spacing $d$,which are design parameters. Next, the SOSP is encoded using the constraints shown in Theorem \ref{SOSPTheorem}. Lastly, as the SOSP is run, the algorithm returns a function, $B(x)$, that is evaluated at any $x_0 \in \mathcal{X}_0$ and utilized to compute the probability, $P$, using (\ref{bound1}), (\ref{bound2}) or (\ref{bound3}). %If $P$ is less than the initialized optimal $P^*$---which is initialized as the trivial value of 1--- then that value is stored as $P^*$ and the algorithm continues its grid search over $\alpha$ to conclusion. Throughout the execution of our algorithm we make an effort to store all relevant values (e.g. $\alpha$, $\beta$, $P$) such that the evolution of the probability bounds can be seen for varying parameters.
	The degree of $B(x)$ is a design parameter; however, higher order polynomials tend to produce tighter bounds. Well refined bounds (i.e. higher order polynomials) present themselves with the trade-off of longer computational times versus probability of failure refinement. \par 
	The objective of the SOSP in Algorithm \ref{Bxalgo} is set to minimize the value, $B(x_0) + \beta$. Minimizing $ B(x_0) + \beta$ is a consensus objective which may not be the best one but provides a means of avoiding bi-linear programs.
        % \begin{remark}
        %   As in Remark \ref{rem:1}, if $x_0$ is not known exactly, $\gamma$ can be substituted for $B(x_0)$ to provide a bound for all initial conditions $x_0\in \mathcal{X}_0$.
        % \end{remark}
        %An alternative objective would be to minimize $\gamma + \beta$.
\begin{algorithm} 
    \footnotesize
	\caption{Compute $B(x)$}
	\begin{algorithmic}[1] 
	\Procedure{Compute-$B$}{$l_{\alpha}, u_{\alpha}, \sigma, u(x), n_B$} 
	\State $\alpha \gets Range(l_{\alpha}, u_{\alpha}, d)$ \Comment Assign $\alpha$ values $d$ apart
	\State $P^* \gets 1$
	\State $P \gets \emptyset$
	\For {$\alpha_0 \in \alpha$}
		\State $\min$ $\ {B(x_0) + \beta}$
		\State $\text{subject to}$  \qquad $B(x) - \lambda_{\mathcal{X}}s_{\mathcal{X}}(x) \geq 0$
		\State  \qquad \qquad  \qquad$-\mathcal{A}B(x)+\alpha_0 B(x)-\beta$
		\State \qquad \qquad \qquad $+\lambda_{\mathcal{X}_u}s_{\mathcal{X}_u}(x) - \lambda_{\mathcal{X}}(x)s_{\mathcal{X}}(x)\geq 0$
		\State  \qquad \qquad \qquad$ - B(x) - \lambda_{\mathcal{X}_o}s_{\mathcal{X}_o}(x) + \gamma \geq 0$ 
		\State  \qquad \qquad \qquad$ B(x) - 1-\lambda_{\mathcal{X}_u}s_{\mathcal{X}_u}(x) \geq 0$
		\State 
		\State Compute $P$, using (\ref{bound1}), (\ref{bound2}) or (\ref{bound3}) %( $\alpha$ and $\beta$ dependent)
		\If{$P < P^*$} 
			\State $\alpha^* = \alpha_0$
			\State $\beta^* = \beta$
			\State $P^* = P$
		\EndIf
	\EndFor
	\State \textbf{return} $\alpha^*, \beta^*, P^*$
	\EndProcedure
	\end{algorithmic}
	\label{Bxalgo}
\end{algorithm}
\setlength{\textfloatsep}{10pt}
    %   \SC[inline]{Should we include a subsection break here? Maybe a subsection for computing $B$ and a subsection for control synthesis}
    %   \CS[inline]{After doing so, it does look like a good point to do a subsection break. Done.}
\subsection{Controller Synthesis Procedure}
In general, when searching for a control policy, we are aiming for a polynomial of the same or lower order of $B(x)$ such that the upper bound on the probability of failure reduces to a designer specified value. First, we write the polynomial $u(x)$ in quadratic form as 
\begin{equation} \label{SMRu}
  u(x) = z^TQz
\end{equation}
where $z$ is a vector of monomials in $x$ of a specified order and $Q$ is a coefficient matrix of appropriate dimensions.  Because there likely exist many feasible controllers ensuring the desired probability of failure, we introduce a cost criterion to choose among them.
%Our goal is to find a control policy that minimizes the required control energy to achieve a desired probability bound, however, this is difficult to incorporate directly in the optimization problem. 
We approximate the energy of a particular control policy via a proxy measure. In this case, the proxy is the non-negative scalar, $c$, such that the following vector element-wise constraints
\begin{equation} \nonumber
 c\mathds{1} - \vect(Q) \geq 0 
\end{equation}
\begin{equation} \nonumber
 \vect(Q) + c\mathds{1}  \geq 0 
\end{equation}
hold where $\vect(Q)$ is the vector form of matrix $Q$ and $\mathds{1}$ is the vector of ones of appropriate dimension. We choose the cost $\min c$ to minimize the coefficients appearing in the polynomial controller to encourage lower control effort. This objective and procedure are highlighted in Algorithm \ref{Uxinit}. 
\begin{algorithm}
    \footnotesize
	\caption{Initialize $u(x)$}
	\begin{algorithmic}[1]
		\Procedure{Compute-$u$}{$B(x), \alpha, \beta, n_u$}
		\State $u(x) = z^TQz$ \Comment{$u(x)$ is an $n_u$ power polynomial}
		\State \Comment{$z$ is a vector of state monomials}
        \State $\min \ c$
        \State subject to \qquad $c\mathds{1} - \vect(Q) \geq 0 $
        \State \qquad \qquad \qquad $\vect(Q) + c\mathds{1}  \geq 0$ 
        \State \qquad \qquad \qquad $-\mathcal{A}B(x) + \alpha B(x) - \beta$ 
        \State \qquad \qquad \qquad  $+\lambda_{\mathcal{X}_u}(x)s_{\mathcal{X}_u}(x) - \lambda_{\mathcal{X}}(x)s_{\mathcal{X}}(x)\geq 0$
		\State \textbf{return} $u(x), c, Q$ 
		\EndProcedure
	\end{algorithmic}
	\label{Uxinit}
\end{algorithm} \par
\setlength{\textfloatsep}{3pt}
Algorithm \ref{Uxalgo} takes $P_{goal},\sigma, \alpha ,n_B, n_u$ and  $\epsilon$ as arguments. These variables are the goal probability, diffusion term, $\alpha$ multiplier on $B(x)$, barrier polynomial order, control polynomial order and a small offset, respectively. Once the procedure begins, it runs until the probability of failure is within $\epsilon$ of the predefined goal probability. It is possible to use other conditions to determine whether the algorithm should continue to run such as computing the change in optimal scalar $c$ value, $c^*$, between iterations. Additionally, the algorithm may be sped up by using a floor value for $c$. Initially, a polynomial barrier of a specified polynomial power is computed given no control policy (i.e. $u(x) = 0$). %Here, we wish to store the probability of failure of the system and the $\beta$ value. If the probability of failure is below the goal probability it is stored.
Generally speaking, as in our case studies, we are interested in systems where the probability of failure with no control action is above the goal probability. \par
Next, if the probability of failure is greater than $P_{goal}$ then we compute a scaled down $\beta$ value multiplied by $a_{dec}$. If the failure probability is less than $P_{goal}$ then we scale up the $\beta$ by $a_{inc}$. The intuition behind this comes from analyzing the probability bounds (\ref{bound1}), (\ref{bound2}) or (\ref{bound3}). In general, a lower $\beta$ reduces our failure probability, thus when searching for $u(x)$ a scaled version of $\beta$ can be used. The values of $a_{inc}$ and $a_{dec}$ are also design parameters. %Algorithm \ref{Uxalgo} runs until a low energy $u(x)$ has been found. 
\par 
\begin{algorithm} \label{binary}
    \footnotesize
	\caption{Search for control polynomial $u(x)$}
	\begin{algorithmic}[1]
		\Procedure{Compute-$u_{min}$}{$P_{goal},\sigma, \alpha ,n_B, n_u, \epsilon$} 
	    \State $i_{count} = 1$ \Comment{Initialize counting variable}
            \While {$|P^* - P_{goal}$ $| > \epsilon $}
		    \If {$i_{count} = 1$}
		    	\State $\beta, P \gets$ {\small COMPUTE-$B$}{$(l_{\alpha}, u_{\alpha}, \sigma, u(x), n_B)$}
		    	\State \Comment{Since $\alpha$ fixed, $l_{\alpha} = u_{\alpha}$}
		    	\State \Comment{$u(x) = 0$}
			    \State $i_{count} = i_{count} + 1$
			\Else 
						    \State $u(x), c, Q \gets$ {\small COMPUTE-$u$}{($B(x), \alpha, \beta, n_u$)}
							\State $\beta, P \gets$ {\small COMPUTE-$B$}{$(l_{\alpha}, u_{\alpha}, \sigma, u(x), n_B)$}
			\EndIf 
			\State 
				\If{$P < P_{goal}$ \textbf{\text{and}} $c < c^*$} 
				\State $\beta^* = \beta$
				\State $P^* = P$
				\State $c^* = c$
				\EndIf
				\State
				\Comment{\small{$c^*$ is initialized as a large number} \normalsize}
				\If {$P > P_{goal}$}
					\State $\beta = a_{dec} \beta$
				\Else 
					\State $\beta = a_{inc} \beta$ 
				\EndIf
				\State 
				\Comment{\small $a_{inc}$ and $a_{dec}$ are scaling factors}
				\State 
		\EndWhile

		\State \textbf{return} $u^*(x), c*, Q$
		\EndProcedure
	\end{algorithmic}
	\label{Uxalgo}
\end{algorithm}
\section{Case Studies}
In this section, we first present a simple academic example to illustrate the our technique. Second, we present a nonlinear example to demonstrate the versatility of our approach. Both case studies are compared to a Monte Carlo simulation which is considered ground truth. We utilize SOSTOOLS \cite{sostools} which converts our SOSP into semidefinite programs. Our choice of solver is the semidefinite program solver SDPT3 \cite{SDTP3_1999, SDTP3}. These case studies were conducted on a 2.3 GHz Intel Core i5 computer with 8GB of memory.\footnote{The \MATLAB source code for the two case studies is contained at \url{https://github.com/gtfactslab/stochasticbarrierfunctions}}  
\subsection{1-D Stochastic System}
 Consider a 1-D stochastic control affine system of the form
\begin{equation}
	dx = \big(-x + u(x)\big)dt + \sigma dw.
\end{equation}
This is of the same form as (\ref{controlstochdiff}) where $f(x) = -x$ and $g(x)  = 1$. We define our state space as $\mathcal{X} = \{x: -2 \leq x \leq 2 \}$, $\mathcal{X}_u = \{x: x^2 \geq 1\}$, and $\mathcal{X}_0 = \{x: x^2 \leq .2^2\}$.
% \SC[inline]{Can we set $\mathcal{X}=\mathbb{R}$?} \CS[inline]{Yes. Done. Is this the best way to write it?} \SC[inline]{Hold on- above, you require $\mathcal{X}$ to be compact. I suppose we should revert to the original version, although I don't know why compactness is required?} \CS[inline]{I took that requirement from Prajna's paper.}
% \SC[inline]{Ok, then I suppose we should use your original state space definition Sorry.}
%When evaluating our probability bounds, we evaluate $B(x_0)$ at $x_0 = 0$ and use this same initial condition in the Monte Carlo simulations. 
First, we benchmark the probability of failure without a control input (i.e. $u(x) = 0$) for a finite time horizon of $T = 1 \ \text{s}$. Thus, to do so, we follow the procedure outlined in Algorithm \ref{Bxalgo}. We grid search over a defined range of values for the constant $\alpha$. In this particular example, our $\alpha \in [0, 5]$ with $d = .05$ in Algorithm \ref{Bxalgo}. \par
\begin{figure} 
    \centering
	\includegraphics[scale=.30]{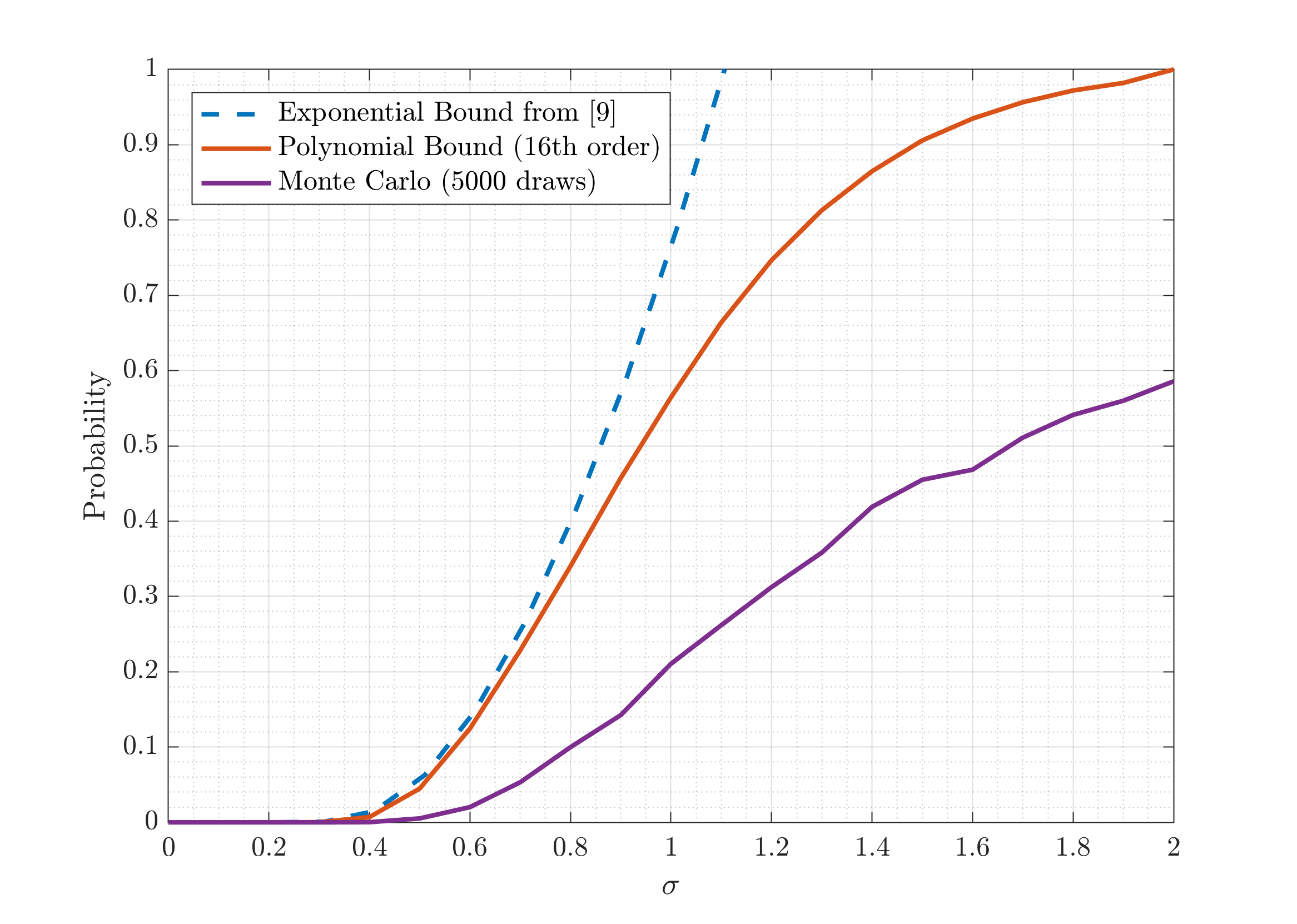}
	\setlength{\belowcaptionskip}{-6pt}
	\caption{The probability of failure bounds for the 1-D system are presented here. The polynomial barrier function, $B(x)$ considered here was of the 16\textsuperscript{th} degree. The Monte Carlo simulation results illustrate the true probability of failure for this system.}
	\label{ex1probbounds}
\end{figure}

 
We choose to search for $B(x)$ of the 16\textsuperscript{th} degree. Additionally, we reproduce the c-martingale bound presented in \cite [Algorithm 3]{SteinhardtJacob2012Frvo}. Lastly, we benchmark against the true probability of failure created via a 5000 draw Monte Carlo simulation. The results are presented in Fig. The  \ref{ex1probbounds}. \par  
In Fig. \ref{ex1probbounds}, we see that our polynomial bound on the probability of failure performs better than the bound from \cite{SteinhardtJacob2012Frvo} generated using the c-martingale condition that is not state dependent. The difference is particularly notable at higher noise levels where the exponential bound from \cite{SteinhardtJacob2012Frvo} becomes trivial, i.e., greater than or equal to one.  \par
Next, we address the control problem of achieving a particular bound on the probability of failure of this system. We choose a desired failure probability $P_{goal} = .30$. We restrict our attention to a linear controller of the form $u(x) = -kx$. Our search for a low-energy controller which successfully fulfills the design requirement follows a modified binary search version of Algorithm \ref{Uxalgo}.% This enables a simple search for the $k$ necessary to achieve our desired criterion.
\begin{figure} 
    \centering
	\includegraphics[scale=.40]{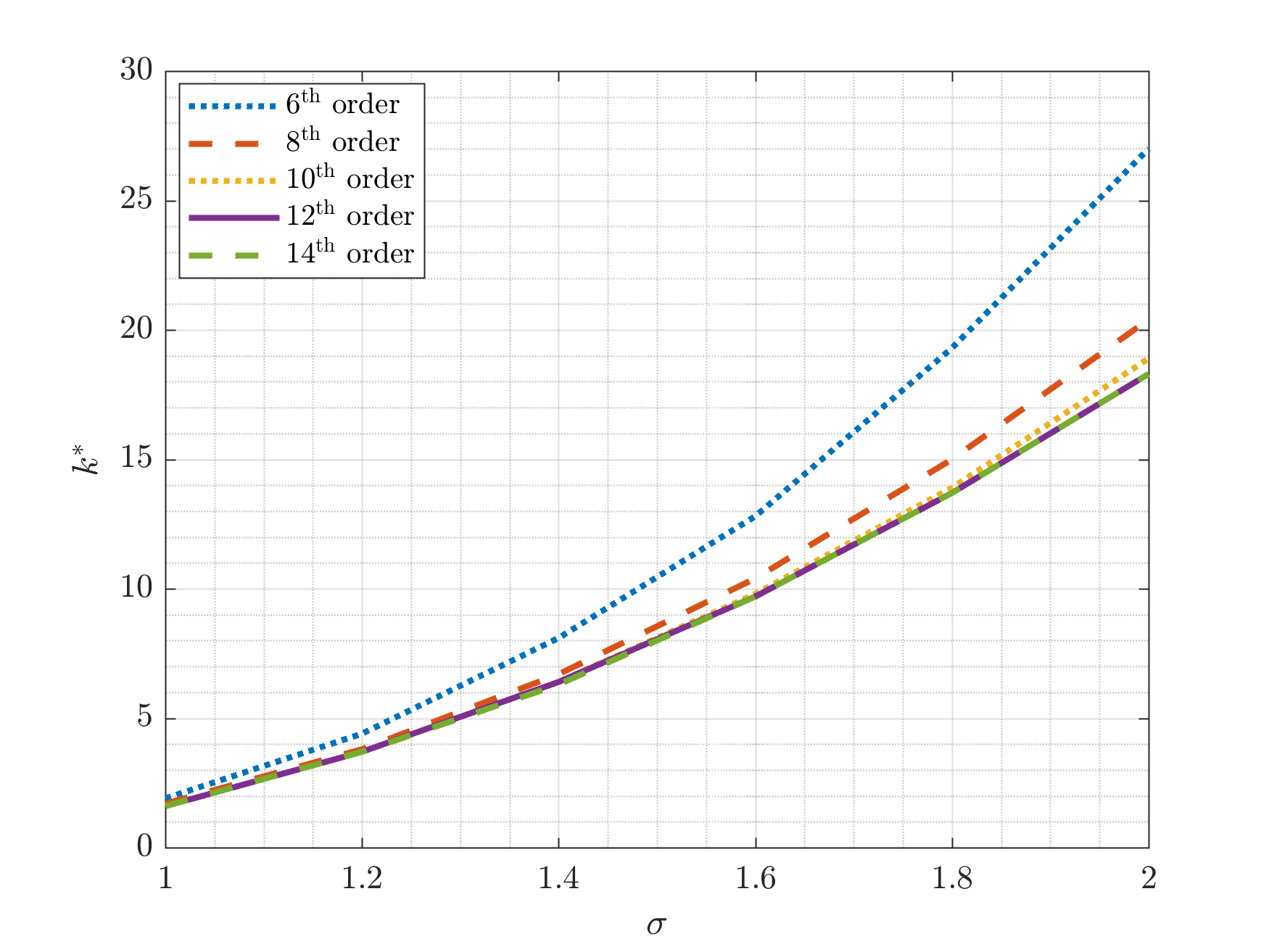}
	\setlength{\belowcaptionskip}{-6pt}
	\caption{ An illustration of the trade-off between required control gain and the degree of the barrier function, $B(x)$ needed to successfully attain the desired probability of failure threshold. Using higher order polynomials allows us to guarantee that the desired probability bound is satisfied for a smaller control gain up until some point. %Eventually, the order of the polynomial will not improve the bound as is beginning to happen from the 12\textsuperscript{th} to 14\textsuperscript{th} order polynomial.}
	}
	\label{ex1controlsearch}
\end{figure}
Fig. \ref{ex1controlsearch} plots $k^*$ achieving the desired failure probability bound for $\sigma \in [1, 2]$. Here, we note that the degree of barrier function for which we search greatly affects the control gain needed to achieve the control objective. In some sense, searching for a higher-order polynomial refines the probability of failure bound requiring lower control effort; however, these high order polynomials require more computation time. Eventually, the degree of the polynomial reaches a saturation point where it does not further decrease the $k^*$ required. 
\subsection{Nonlinear Dynamics}
Consider the stochastic non-linear dynamics % from \cite{Prajna:2004up} with an additional input term $u(x)$:
\begin{align}
    dx_1 &= x_2dt \label{vdp1}  \\
    dx_2 &= \bigg(-x_1 - x_2 - 0.5x_1^3+ u(x) \bigg) dt+ \sigma dw.
    \label{vdp2}
\end{align}
This system is studied in \cite{Prajna:2004up} without the input term $u(x)$.
\par
\begin{figure} 
	\centering
	\includegraphics[scale=.30]{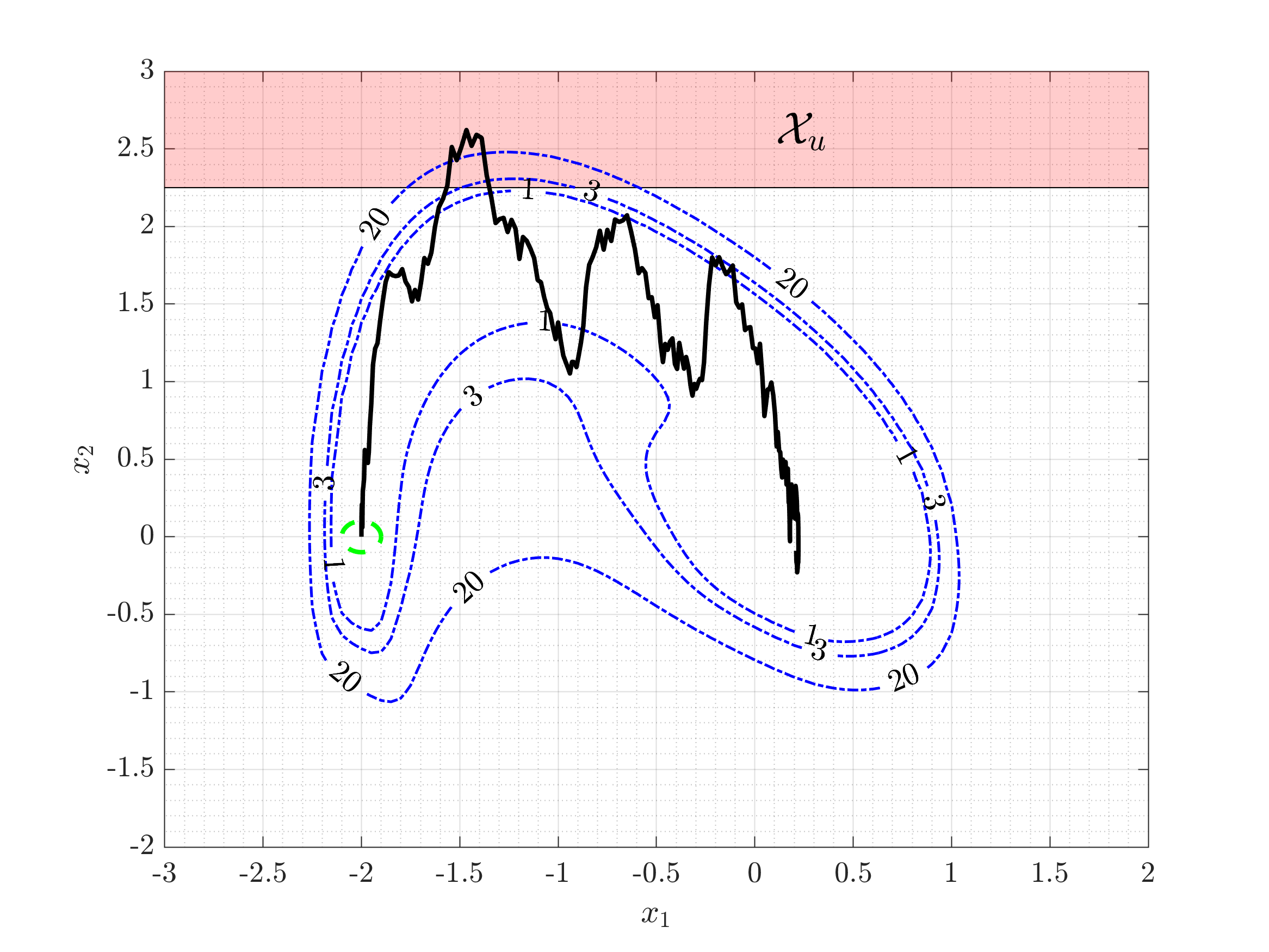}
	\setlength{\belowcaptionskip}{-10pt}
	\caption{Given the initial conditions $x_0 = [-2, 0]$, the single trajectory dynamics for time horizon of $T = 2$ and a $\sigma = 1.0$ is illustrated. We define the unsafe region as $\mathcal{X}_u = \{x_2 \ | \ x_2 \geq 2.25\}$. Additionally, the level sets of $B(x)$ and their respective values are labeled and given as dashed blue lines. }
	\label{ex2problemillustration}
\end{figure}
\begin{figure} 
	\centering
	\includegraphics[scale=.30]{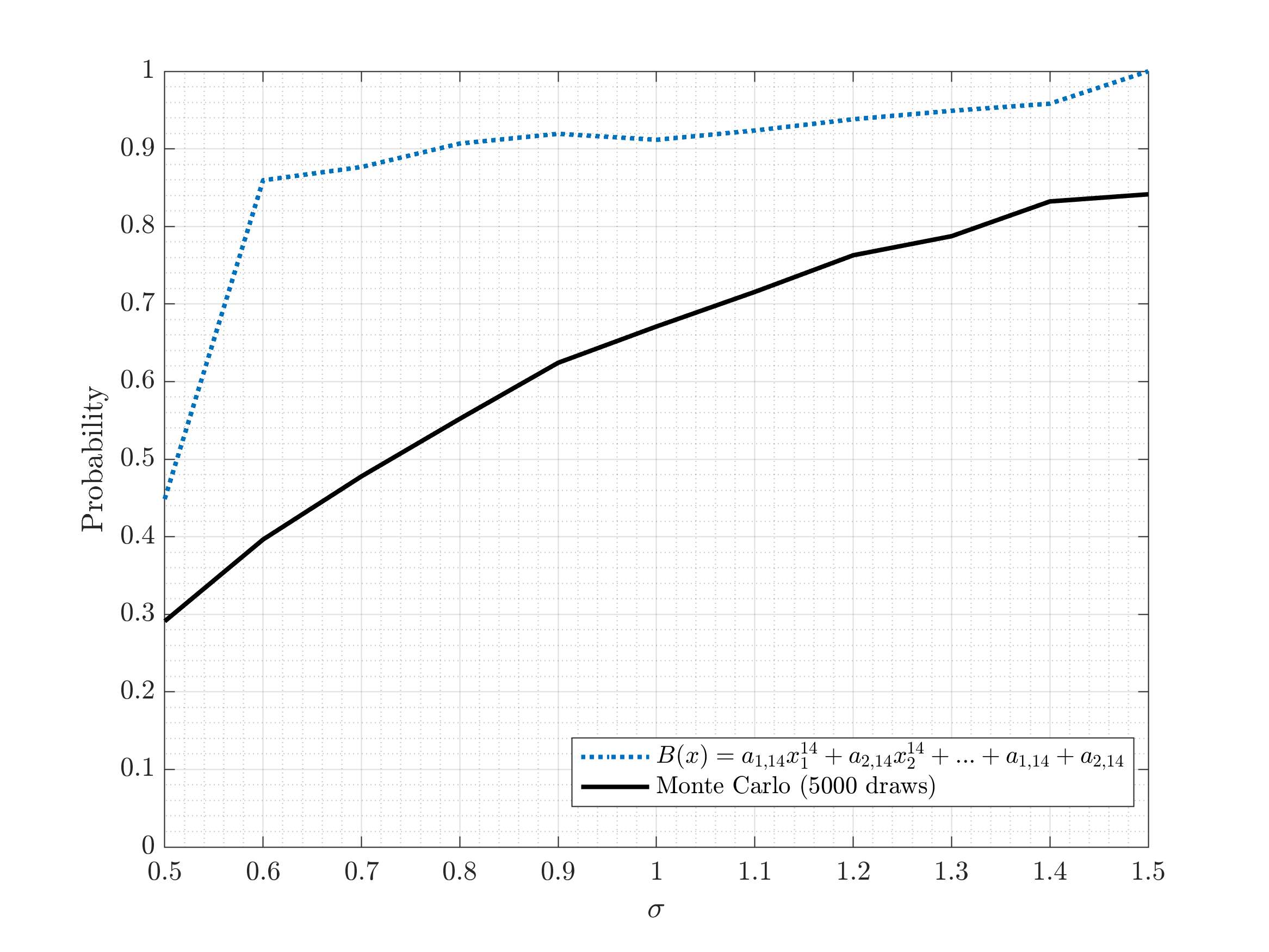}
	\setlength{\belowcaptionskip}{-2pt}
	\caption{Computing a 14\textsuperscript{th} order polynomial barrier function for the nonlinear dynamics we are able to bound the probability of failure of the 5000 draw Monte Carlo dynamics for constant noise levels $\sigma \in [.5, 1.5]$.}
	\label{ex2probabilitybound}
\end{figure}

 We define our state space as $\mathcal{X} = \{(x_1, x_2) \ | -3 \leq x_1 \leq 2, -2 \leq x_2 \leq 3\}$, $\mathcal{X}_u = \{x_2 \ | \ x_2 \geq 2.25 \}$, and $\mathcal{X}_0 = \{(x_1, x_2)| (x_1 + 2)^2 + x_2^2 \leq 0.1^2 \}$. A sample trajectory of (\ref{vdp1})--(\ref{vdp2}) is illustrated in Fig. \ref{ex2problemillustration}. Additionally, level sets of $B(x)$ are projected onto the state space. In this illustration, $B(x)$ is computed with $u(x) = 0$ solely using Algorithm \ref{Bxalgo}. Here, we see that the values for the barrier function abide by the definitions of Theorem \ref{alphatheorem}. In this particular trajectory illustration, the evolution of system noise is enough for the system to enter the predefined unsafe set; however, this is not always the case. To illustrate this, we compute a Monte Carlo simulation of the system dynamics shown. Additionally, an upper bound is computed on the probability of becoming unsafe given our initial condition and illustrated in Fig. \ref{ex2probabilitybound}. While we encode a set of initial conditions into the SOSP, we evaluate the probability bound at the same initial point, $x_0 \in \mathcal{X}_0$, as the Monte Carlo simulation. \par
%We now address the problem of finding a state feedback control strategy to achieve a desired safety metric. 
The design specification for this example is to reduce the probability of failure bound to $P_{goal} = .10$ for specified noise levels. For this example we create a 2\textsuperscript{nd} order polynomial controller of the form of (\ref{SMRu}). We look to minimize the constant, $c$, highlighted in Algorithm \ref{Uxinit}. We run the $u(x)$ search algorithm for select noise levels, specified $\alpha$ values, and present the results in Table \ref{controlresult}. The $\alpha$ values in this table originate from the initial (i.e. $u(x) = 0$) probability bound computation. %Here, we consider 10\textsuperscript{th} order $B(x)$ due to computational limitations of SOSTOOLS.
\begin{table}[H]  
    \scriptsize
    \centering 
    \def\arraystretch{1.35}
     \begin{tabular}    {|| >{\centering\arraybackslash}m{.70in} | >{\centering\arraybackslash}m{.58in}| >{\centering\arraybackslash}m{.58in}| >{\centering\arraybackslash}m{.68in} ||}
        \hline
         \textbf{Noise Level}, $\mathbf{\sigma}$ & $\mathbf{P_{u(x) = 0}}$ & $\large{\mathbf{\alpha}}$ &$ \mathbf{\min c}$\\ [0.1ex] 
         \hline\hline
         0.6 & 0.860 & 1.4 & 2.1821\\
         \hline
         0.9 & 0.919 & 1.3 & 0.5251\\
         \hline
         1.0 & 0.912 & 1.3 & 0.6396\\
         \hline
         1.3 & 0.949 & 1.5 & 1.1488 \\
         \hline
    \end{tabular}
	\setlength{\belowcaptionskip}{-2pt}
    \caption{The results from the search for a control polynomial u(x) which reduces the probability of failure to $P_{goal} = .10$. The probability of failure without a given control input is presented here for comparison.}\label{controlresult}
\end{table}
\section{Conclusion}
We consider control barrier functions whose existence gives a means of quantifying an upper bound on a system's probability of failure. Additionally, we present a novel, state dependent approach to the problem of finite-time verification which further relaxes the constraint on the evolution of the expected value. Lastly, we synthesize a feedback control strategy $u(x)$ such that a certain probability of failure criteria is met. We illustrate our methods with two case studies which demonstrate our ability to quantify system failure probabilities. In these case studies, we solve for the barrier function polynomials using SOS optimization and demonstrate our proposed approach outperforms existing methods.

\bibliographystyle{IEEEtran}
\bibliography{references.bib}
\end{document}